\documentclass[a4paper,fleqn,usenatbib]{mnras}
\usepackage{newtxtext,newtxmath}
\usepackage{cite}
\usepackage{setspace}

\usepackage[T1]{fontenc}
\usepackage{ae,aecompl}

\usepackage[dvipdfmx]{graphicx}
\usepackage{amsmath}	
\usepackage{amssymb}	
\usepackage{comment}
\usepackage{color}

\title[An Estimation of the White Dwarf Mass in the Dwarf Nova GK Persei]{An Estimation of the White Dwarf Mass in the Dwarf Nova\\GK Persei with NuSTAR Observations of Two States}

\author[Y. Wada et al.]{
Yuuki Wada,$^{1,2}$\thanks{E-mail: wada@juno.phys.s.u-tokyo.ac.jp}
Takayuki Yuasa,$^{3}$
Kazuhiro Nakazawa,$^{1}$
Kazuo Makishima,$^{4}$\newauthor
Takayuki Hayashi,$^{5,6}$
and Manabu Ishida$^{7,8}$
\\
$^{1}$Department of Physics, Graduate School of Science, The University of Tokyo, 7-3-1 Hongo, Bunkyo-ku, Tokyo 113-0033, Japan\\
$^{2}$High Energy Astrophysics Labolatory, Nishina Center for Accelerator-Based Science, RIKEN, 2-1 Hirosawa, Wako, Saitama 351-0198, Japan\\
$^{3}$55 Devonshire Road 239855, Singapore\\
$^{4}$MAXI team, RIKEN, 2-1 Hirosawa, Wako, Saitama 351-0198, Japan\\
$^{5}$Department of Graduate School of Science, Nagoya University, Furo-Cho, Chikusa-ku, Nagoya 464-8602, Japan\\
$^{6}$Goddard Space Flight Center, National Aeronautics and Space Administration, Greenbelt, MD 20771, USA\\
$^{7}$Institute of Space and Astronautical Science, Japan Aerospace Exploration Agency, 3-1-1 Yoshinodai, Chuo-ku, Sagamihara 252-5210, Japan\\
$^{8}$Department of Physics, Tokyo Metropolitan University, 1-1 Minami-Osawa, Hachioji, Tokyo 192-0397, Japan
}

\date{Accepted 2017 November 03. Received 2017 October 22; in original form 2017 September 05}

\pubyear{2017}

\begin{document}
\label{firstpage}
\pagerange{\pageref{firstpage}--\pageref{lastpage}}
\maketitle
\begin{abstract}
We report on X-ray observations of the Dwarf Nova GK Persei performed by {\it NuSTAR} in 2015.
	GK Persei, behaving also as an Intermediate Polar, exhibited a Dwarf Nova outburst in 2015 March--April.
	The object was observed with {\sl NuSTAR} during the outburst state, and again in a quiescent state wherein the 15--50 keV flux was 33 times lower.
	Using a multi-temperature plasma emission and reflection model,
	the highest plasma temperature in the accretion column was measured as $19.7^{+1.3}_{-1.0}$~keV in outburst and $36.2^{+3.5}_{-3.2}$~keV in quiescence.
	The significant change of the maximum temperature is considered to reflect an accretion-induced decrease of the inner-disk radius $R_{\rm in}$,
	where accreting gas is captured by the magnetosphere.
	Assuming this radius scales as $R_{\rm in} \propto \dot{M}^{-2/7}$ where $\dot{M}$ is the mass accretion rate,
	we obtain $R_{\rm in} = 1.9 ^{+0.4}_{-0.2}~R_{\rm WD}$ and $R_{\rm in} = 7.4^{+2.1}_{-1.2}~R_{\rm WD}$ in outburst and quiescence respectively, 
	where $R_{\rm WD}$ is the white-dwarf radius of this system.
	Utilising the measured temperatures and fluxes, as well as the standard mass-radius relation of white dwarfs, 
	we estimate the white-dwarf mass as $M_{\rm WD} = 0.87~\pm~0.08~M_{\rm \odot}$ including typical systematic uncertainties by 7\%.
	The surface magnetic field is also measured as $B \sim 5 \times 10^{5}$~G.
	These results exemplify a new X-ray method of estimating $M_{\rm WD}$ and $B$ of white dwarfs by using large changes in $\dot{M}$.
\end{abstract}




\section{Introduction}
\label{sec:introduction}
\begin{figure*}
	\begin{center}
	\includegraphics[width=0.7\textwidth]{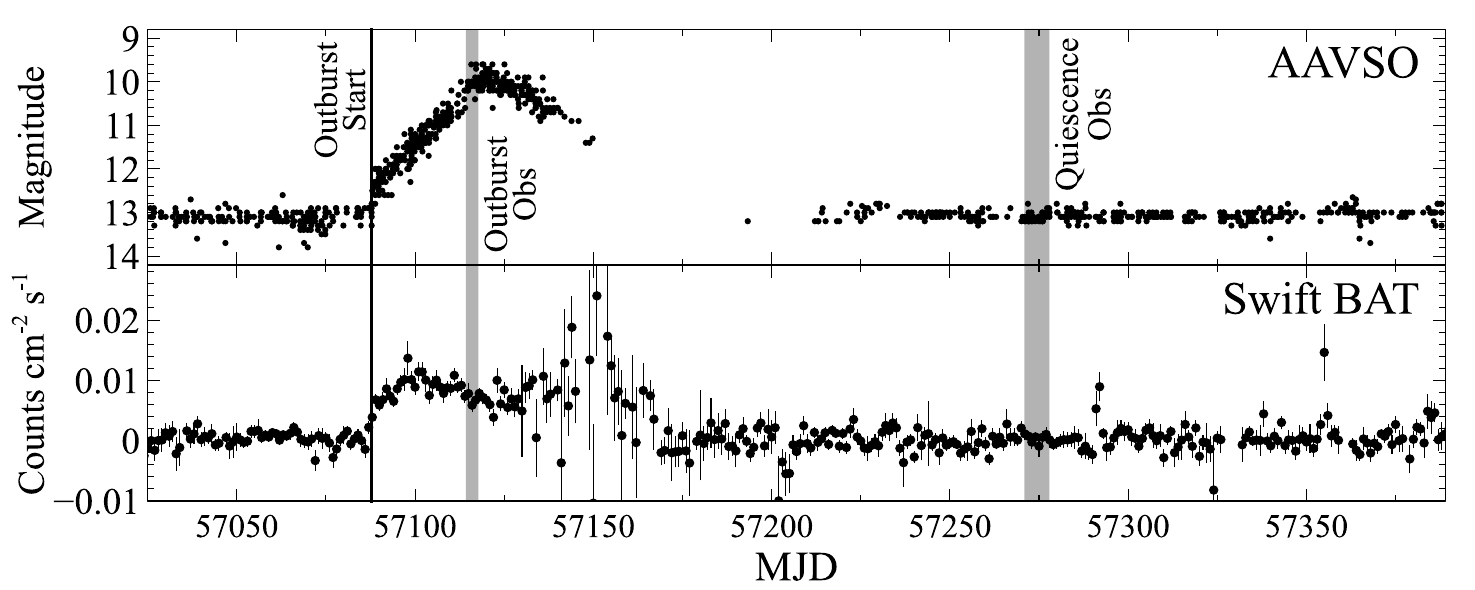}
	\caption{Optical light curves of GK Persei without a filter from {\sl AAVSO International Database}, 
	and the {\sl Swift}/BAT \citep{2013ApJS..209...14K} 15--150 keV count rate history from {\sl Swift}/BAT X-ray Transient Monitor web site.
	The shaded regions indicate the two {\sl NuSTAR} observations.}
	\label{fig:monitor}
	\end{center}
\end{figure*}

\begin{table*}
	\begin{center}
	\caption{The present observation log of GK Persei by {\sl NuSTAR}.}
	\begin{tabular}{cccccc}
		\hline
					& Observation ID	& Start Date/Time 		& Stop Date/Time 		& Exposure$^{\rm a}$	& Count rate$^{\rm b}$	\\ \hline
		Outburst 		& 90001008002	& 2015-04-04 02:46:07	& 2015-04-06 15:10:35	& 42					& 18.09 $\pm$ 0.02		\\
		Quiescence 	& 30101021002	& 2015-09-08 15:46:08	& 2015-09-11 02:04:09	& 72					& 1.080 $\pm$ 0.006		\\ \hline
	\multicolumn{6}{l}{$^{\rm a}$ A net exposure of each of FPMA and FPMB in ks.}\\
	\multicolumn{6}{l}{$^{\rm b}$ Averaged 3--50 keV combined count rates of FPMA plus FPMB in units of count s$^{-1}$.}\\
	\end{tabular}
	\label{table:log}  
	\end{center}
\end{table*}

Cataclysmic Variables (CVs) are close binary systems consisting of a mass-accreting white-dwarf (WD) primary and a mass-donating companion. 
	Gas overflowing from the Roche lobe of the companion accretes onto the WD surface, where gravitational energy of the gas is converted mainly into X-ray emission. 
	CVs hosting a magnetised WD are further classified into ``Polars'' and ``Intermediate Polars'' (IPs),
	in which the WDs have magnetic-field strengths of $B \sim 10^{7-9}$~G and $B \sim 10^{5-7}$~G, respectively. 

In an IP, the gas from the companion forms an accretion disk down to a radius $R_{\rm in}$ where the gravity working on the accreting matter is counter-balanced by the magnetic pressure. 
	Then, the gas is captured by the WD's magnetosphere, and accretes onto the WD surface to form a pair of accretion columns due to the strong magnetic field.
	In the accretion columns, the gas is heated to $10^{7-8}$~K by a standing shock, and lands onto the WD surface after releasing most of its energies into thermal X-rays.
	If $R_{\rm in}$ is far enough from the WD surface, the temperature $T_{\rm s}$ just below the shock is proportional to the gravitational potential of the WD \citep{1973PThPh..49.1184A} as
	\begin{equation}
		kT_{\rm s} = \frac{3}{8} \mu m_{\rm p} \frac{GM_{\rm WD}}{R_{\rm WD}},
		\label{eq:aizu}
	\end{equation}
	where $\mu$ is the mean molecular weight, $m_{\rm p}$ is the proton mass, $M_{\rm WD}$ is the WD mass, and $R_{\rm WD}$ is its radius. 
	Therefore, $M_{\rm WD}$ can be estimated by combining the measured $T_{\rm s}$ with the standard mass v.s. radius ($M_{\rm WD}$-$R_{\rm WD}$) relation of WDs \citep{1972ApJ...175..417N}
	\begin{equation}
		R_{\rm WD} = 7.8 \times 10^{8}~ {\rm cm} \left[ \left( \frac{1.44 M_{\odot}}{M_{\rm WD}}\right)^{2/3} - \left( \frac{M_{\rm WD}}{1.44 M_{\odot}}\right)^{2/3} \right]^{1/2}.
		\label{eq:nauenberg}
	\end{equation}

An X-ray spectrum from an IP is a particular superposition of optically-thin thermal emissions of various temperatures, from $T_{\rm s}$ downwards \citep{1998MNRAS.293..222C}.
	To determine $T_{\rm s}$, it is hence important to accurately measure both the hard X-ray continuum (e.g. \citealt{2005A&A...435..191S,2010A&A...520A..25Y}),
	and the ratio of Fe XXV and XXVI lines at $\sim 7$ keV \citep{1997ApJ...474..774F}. 
	This is because the former is sensitive to the hottest components (with temperature $\sim T_{\rm s}$),
	whereas the latter tells us contributions from cooler components arising closer to the WD surface. 

GK Persei, at an estimated distance of $477 ^{+28}_{-25}$ pc \citep{2013ApJ...767....7H}, interestingly exhibits three distinct aspects of CVs; it behaves as an IP, as a Dwarf Nova,
	and exhibited a classical Nova explosion in 1901 \citep{1901MNRAS..61..337W,1901ApJ....13..173H}.
	It repeats Dwarf Nova outbursts every 2--3 years, each lasting for 2 months (e.g. \citealt{2002A&A...382..910S}). During outbursts, the optical and X-ray luminosities both increase by a factor of 10--20. 
	
By optical observations, \citet{1994A&A...281..108R} and \citet{2002MNRAS.329..597M} obtained lower limits of the WD mass in GK Persei
	as $M_{\rm WD} \geq 0.78 M_{\rm \odot}$ and $M_{\rm WD} \geq 0.55 M_{\rm \odot}$ respectively, and upper limits of the inclination angle as $i \leq 73^{\rm \circ}$ due to lack of eclipses. 
	Through a model fitting to the Nova outburst light curve observed in 1901, \citet{2007ApJ...662..552H} also derived $M_{\rm WD} = 1.15 \pm 0.05 M_{\odot}$. 
	\citet{1999ApJS..120..277E} and \citet{2005A&A...435..191S} measured the shock temperature in outbursts, and derived $M_{\rm WD} = 0.52 ^{+0.34}_{-0.16} M_{\odot}$ with {\sl ASCA} and
	$M_{\rm WD} = 0.59 \pm 0.05 M_{\odot}$ with {\sl RXTE}, respectively.
	However, \citet{2005A&A...435..191S} pointed out that the WD mass based on the outburst observation could be underestimated by at least 20\%.	
	In fact, accretion onto the WD can occur only if $R_{\rm in}$ is smaller than the co-rotation radius, defined as
	\begin{equation}
		\frac{R_{\rm \Omega}}{R_{\rm WD}} = 2.3 \left( \frac{P}{1~{\rm min}} \right)^{2/3} \left( \frac{M_{\rm WD}}{M_{\rm \odot}} \right)^{1/3} \left( \frac{R_{\rm WD}}{10^{9}~{\rm cm}} \right)^{-1}
		\label{eq:corotation}
	\end{equation}
	(e.g. \citealt{1995CAS....28.....W}), where the Keplerian rotation period is equal to the spin period $P$.
	When $P=351~{\rm sec}$ of GK Persei (e.g. \citealt{1985MNRAS.212..917W}) and $M_{\rm WD} \sim 0.8~M_{\rm \odot}$ are employed, 
	$R_{\rm in} < R_{\rm \Omega} \sim 10~R_{\rm WD}$ should be required even in quiescence.
	Therefore,  the condition $R_{\rm in} \gg R_{\rm WD}$ may not generally hold, particularly in outbursts.
	Actually, \citet{2009A&A...496..121B} utilised the {\sl Swift}/BAT survey data during quiescence and obtained $M_{\rm WD} = 0.90 \pm 0.12~M_{\rm \odot}$,
	which is higher than the estimates from the past X-ray results in outbursts.

A recent outburst from GK Persei started in March 2015, and continued for 2 months \citep{2015ATel.7217....1W}.
	During this outburst, \citet{2017MNRAS.469..476Z} triggered a Target of Opportunity (ToO) observation with {\sl NuSTAR} and measured a high spin modulation even in a hard X-ray range.
	\citet{2016A&A...591A..35S} also analysed the ToO data and constrained the WD mass as $M_{\rm WD} = 0.86 \pm 0.02 M_{\rm \odot}$.
	The onset of this outburst was serendipitously caught by {\it Suzaku}, and the obtained data allowed \citet{2016MNRAS.459..779Y} to study the accretion geometry at the beginning of the outburst.

With {\sl NuSTAR}, we observed GK Persei again, after the object returned to its quiescence. Although previous observations of the object in quiescence were unable to detect the hard X-ray component,
	the high sensitivity of {\sl NuSTAR} has for the first time allowed us to detect its hard X-rays (typically in energies above $\sim$20 keV) in quiescence.
	The present paper describes a combined analysis of the outburst and quiescence data from {\sl NuSTAR}, 
	and presents a new method to determine $R_{\rm in}$, $M_{\rm WD}$, and $B$ of the WD in GK Persei utilising the large change in $\dot{M}$.

\section{Observations and data reduction}
\label{sec:observation}
The 2015 outburst of GK Persei started on 2015 march 6.84 UT \citep{2015ATel.7217....1W}.
	As shown in Figure~\ref{fig:monitor}, the ToO observation with {\sl NuSTAR} \citep{2013ApJ...770..103H} was conducted in the middle of the outburst from 2015 April 4 02:46:07 to April 6 15:10:35. 
	The net exposures of FPMA and FPMB are 42~ks each.
	The second {\sl NuSTAR} observation, in quiescence, was performed from 2015 September 8 15:46:08 to September 11 02:04:09 with a net exposures of 72~ks.
	The log of the two observations is given in Table~\ref{table:log}.
	
We utilised the data analysis software package {\tt HEASOFT} version 6.20 and a detector calibration database {\tt NuSTAR CALDB} version 20170222,
	both released and maintained by HEASARC at NASA Goddard Space Flight Center. 
	Photon events in the data sets were extracted with an exclusive data reduction software for {\sl NuSTAR} ``{\tt nupipleine}'' version 0.4.6 and ``{\tt nuproducts}'' version 0.3.0. 
	The on-source events were accumulated from a circular region with a radius of $150''$ (in outburst) and $80''$ (in quiescence) centred on the source.
	The background data were accumulated over a region outside a circle of radius of $170''$ and $100''$ in outburst and quiescence, respectively. 
	The X-ray spectra were analysed and fitted with {\tt XSPEC} version 12.9.1 \citep{1996ASPC..101...17A}.

Generally, the X-ray emission from an IP is pulsed at its $P$.
	In fact, pulsations of GK Persei in the X-ray band have been detected at $P=351~{\rm sec}$ both in outbursts and quiescence
	(e.g. \citealt{1985MNRAS.212..917W,1988MNRAS.231..783N,1992MNRAS.254..647I}).
	In the 2015 outburst observation, \citet{2017MNRAS.469..476Z} clearly detected the 351~sec pulsation both in the 3--10~keV and 10--79~keV ranges.
	In the quiescence observation by {\sl NuSTAR}, we detected a faint pulsation with a modulation amplitude of $\sim$~10\% in the 3--50~keV range.
	In the present paper, we concentrate on spectral analysis and postpone the study of this pulsation for the next publication. 

\section{Analysis and Results}
\label{sec:result}
\begin{figure}
	\begin{center}
	\includegraphics[width=0.4\textwidth]{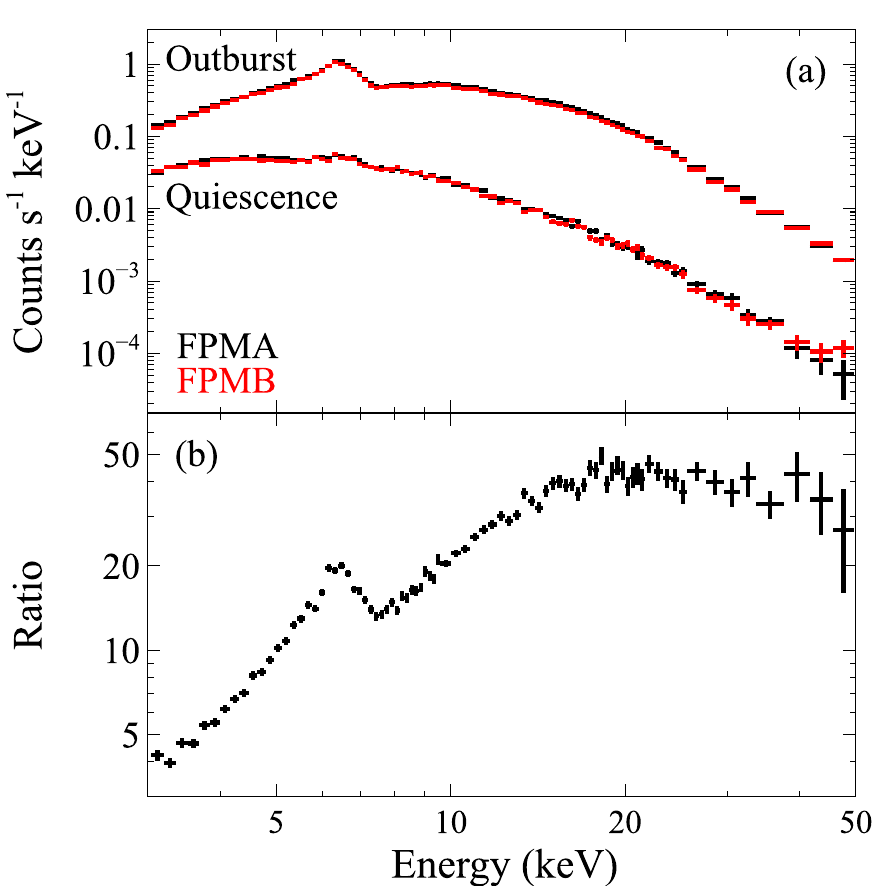}
	\caption{(a) Time averaged FPMA (black) and FPMB (red) spectra of the outburst (brighter) and quiescence (dimmer) observations.
		(b) Ratios of the FPMA plus FPMB spectra between the two observations.}
	\label{fig:ratio}
	\end{center}
\end{figure}

Figure~\ref{fig:ratio} shows 3--50 keV spectra of the outburst and quiescence observations. The background has been subtracted, but the instrumental response has not been removed. 
	Data of FPMA and FPMB are separately plotted. As reported by \citet{2017MNRAS.469..476Z} and \citet{2016A&A...591A..35S}, the hard X-ray continuum is detected up to 70 keV during the outburst. 
	In quiescence, the source is detected up to 50~keV for the first time.
	The 3-50~keV count rate of FPFA plus FPMB was 18.09 $\pm$ 0.02 count s$^{-1}$ in outburst, and 1.080 $\pm$ 0.006 count s$^{-1}$ in quiescence. 
	Thus, the outburst data have 17.5 times higher count rate than those in quiescence. The spectra, particularly the outburst data, exhibit Fe-K line complex at $\sim$ 6.4 keV. 
	From this energy, we regard the lines as mainly of fluorescence origin (from the WD surface and/or the accreting cold matter), rather than ionised lines from the accretion columns.

In the bottom panel, we show the ratio between the two spectra. It reveals three features of the outburst spectrum, in comparison with that in quiescence. 
	Namely, a stronger low-energy absorption, the stronger Fe-K line, and a continuum break at $\sim$ 20 keV.

To analyse the spectra, we employed a multi-temperature optically-thin plasma model {\tt cemekl} \citep{1997MNRAS.288..649D} based on a thermal plasma code {\tt mekal}
	\citep{1985A&AS...62..197M,1986A&AS...65..511M,1995ApJ...438L.115L,1996uxsa.conf..411K}. 
	The differential emission measure of {\tt cemekl} is proportional to the power law function of the plasma temperature $T$ as
	\begin{equation}
	d({\rm EM}) \propto \left(T/T_{\rm s} \right)^{\alpha -1} dT,
	\label{eq:emission}
	\end{equation}
	where $\alpha$ is a positive parameter.
	When the accretion column has a cylindrical shape, $\alpha$ is theoretically calculated as 0.43 by \citet{2005A&A...444..561F}, who used the spectral model computed by \citet{2005A&A...435..191S}.
	We employed this value because high accretion rate systems such as GK Persei are considered to have nearly cylindrical accretion columns \citep{2014MNRAS.438.2267H,2014MNRAS.441.3718H}.
	To imitate the reflection effect on the WD surface, {\tt reflect} model \citep{1995MNRAS.273..837M} was utilised.
	The solid angle of reflector from the irradiator was set to $2\pi$ assuming that the standing shock is formed near the WD surface. 
	The abundances of the {\tt cemekl} and {\tt reflect} components were constrained to be the same assuming that the WD surface near the accretion column is covered by accreted material.
	A gaussian emission model was also added to represent Fe-K line.

\begin{figure*}
	\includegraphics[width=0.8\textwidth]{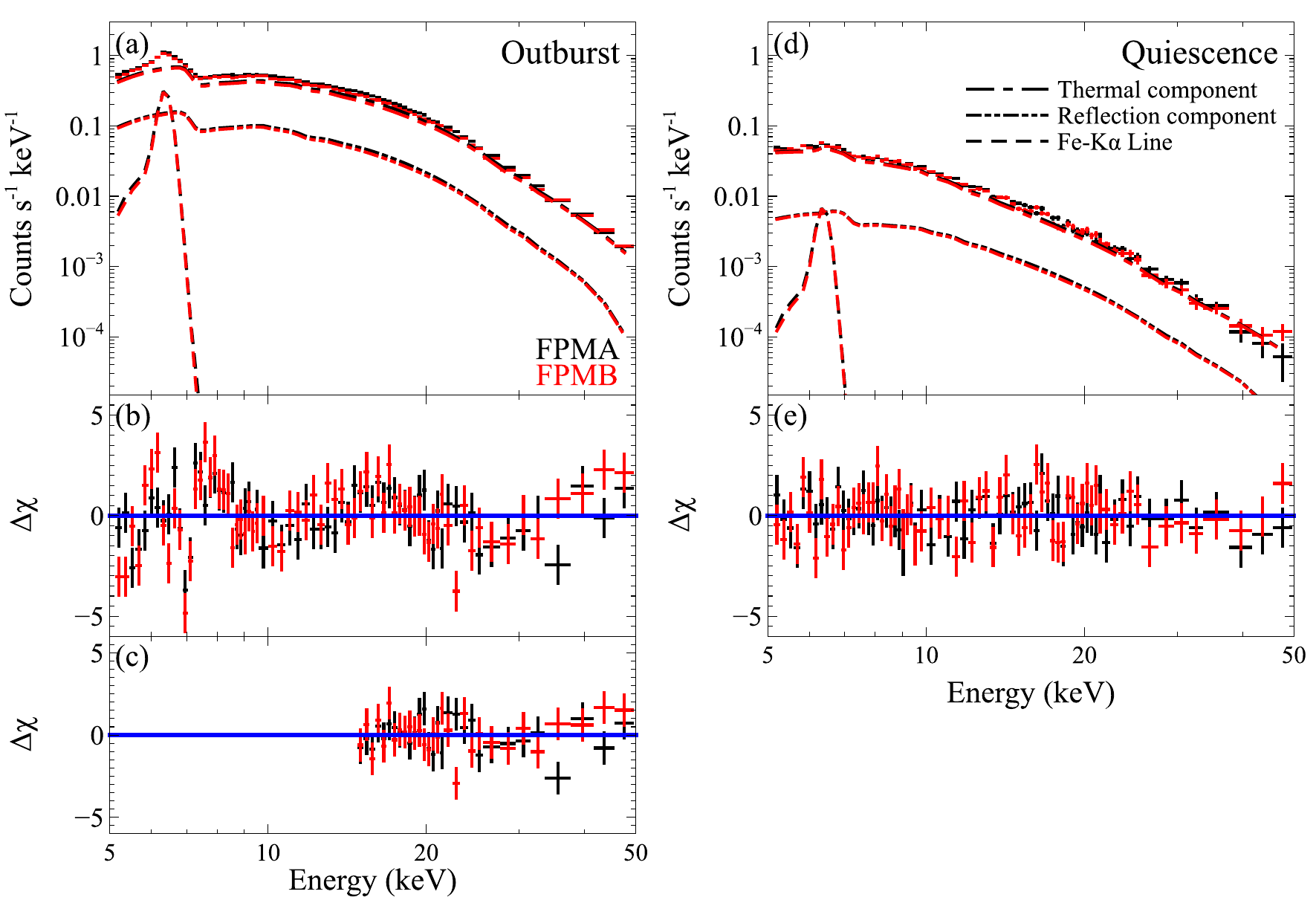}
	\caption{(a) The {\sl NuSTAR} FPMA (black) and FPMB (red) spectra in outburst, and their best-fit models with a partial covering absorption. 
			The thermal component, reflection component, and Fe-K line are indicated by the dash-dot, dash-dot-dot, and dashed lines, respectively. (b) The residuals from the 5--50 keV fit in panel (a).
			(c) The residuals of the 15--50 keV fit. (d) The quiescence spectrum and its best-fit model. (e) The residuals of the fit in panel (d).}
	\label{fig:fitting}
\end{figure*}

With the model thus constructed, we first fitted the outburst spectrum in the 5--50 keV range, because the {\tt cemekl} model is not available above 50~keV.
	A partial covering absorption model was applied to the spectral model in addition to a single column absorber.
	As shown in Figure~\ref{fig:fitting}a, this model approximately reproduced the spectrum, but the fit was formally not acceptable under a 90\% confidence level,
	with the reduced chi-squared of $\chi^{2}/ \nu = 2.26$ for 133 degrees of freedom even including 1\% systematic error.
	In fact, as shown in Figure~\ref{fig:fitting}b, significant residuals were seen in the low energy band ($<$10~keV).

The above fit failure to the outburst spectra is not surprising, because X-ray spectra of IPs are often subject to strong and complex absorption
	that is not modeled by partial absorption (e.g. \citealt{1999ApJS..120..277E}).
	Since refining the absorption model is beyond the scope of the present paper due to lack of constraining data other than the continuum shape,
	we have resorted to discarding low-energy ranges until the effects of complex absorption become negligible 
	(see Ezuka \& Ishida 1999 for a similar method utilised to avoid complex absorption from affecting the spectral fitting result).
	By limiting the fit range to 15--50 keV, the fit to the outburst spectrum has become acceptable even with a single column density absorption.	
	The range of $T_{\rm s}$ constrained in this way, $T_{\rm s} = 19.4 \pm 0.8$~keV, 
	approximately accommodates the value of $T_{\rm s} = 17.9$~keV obtained using the 5--50~keV range (though the fit was unacceptable). 
	
The 5--50~keV quiescence spectrum has been reproduced successfully by the spectral model with a single column absorber.
	Therefore, we have finally conducted a simultaneous fitting using the 15--50~keV band of the outburst spectrum and the 5--50~keV band of the quiescence spectrum.
	The inclination angle and the abundance were set in common, while the other parameters were allowed to vary independently.
	This fitting including 1\% systematic error has become acceptable with $\chi^{2}/ \nu = 1.01$ for 191 degrees of freedom.
	The fit result and the best-fit parameters are presented in Figure~\ref{fig:fitting} and Table~\ref{table:parameter}, respectively. Errors are at 90\% confidence level. 
	The shock temperature was constrained as $T_{\rm s} = 19.7^{+1.3}_{-1.0}$ keV in outburst, and $T_{\rm s} = 36.2^{+3.5}_{-3.2}$ keV in quiescence;
	thus, $T_{\rm s}$ was significantly higher in the latter. The 15-50 keV absorbed flux in outburst was 33 times higher than that in quiescence.
	The 15--50~keV luminosity is also derived as $1.2^{+0.2}_{-0.3} \times 10^{34}$~erg~s$^{-1}$ in outburst, and $3.5^{+0.3}_{-0.5} \times 10^{32}$~erg~s$^{-1}$ in quiescence, 
	with the estimated distance of $477$~pc \citep{2013ApJ...767....7H}.

For our purpose, we need to calculate the total X-ray flux $F$ which is thought to represent the gravitational energy release from $R_{\rm in}$ to $R_{\rm WD}$.
	Starting from the absorbed 15--50 keV flux, $F$ was derived in the following way.
	First, the absorption and the reflection were removed. 
	Second, the flux above 50 keV was included by extrapolating the best-fit model up to 100 keV. 
	Finally, the contribution below 15 keV was incorporated by integrating the best-fit model down to 0.01 keV. The flux above 100 keV and below 0.01 keV are both estimated to be $\ll 0.01 F$. 
	The difference in $F$ between the two spectra amounts to a factor of 65.
	
Since our final fit to the outburst spectrum was obtained by discarding the data below 15~keV, obviously no information was obtained on the Fe-K line.
	Hence we fitted the 5--9~keV spectra of the outburst and quiescence to constrain the Fe-K line equivalent width (EW).
	With a simple model of a single-temperature bremsstrahlung, a gaussian, and a single column absorption model, 
	the EW was measured to be $192 \pm 13$~eV and $52^{+34}_{-26}$~eV in outburst and quiescence respectively, as presented in Table~\ref{table:parameter}.
	The latter is consistent with the value obtained by the 5-50~keV simultaneous fitting.
	
\begin{table*}
\begin{center}
\caption{Best-fit parameters of the multi temperature plasma emission and reflection model to the time-averaged spectra.}
\label{table:parameter}
\begin{tabular}{cccccccccc}
\hline
			& $N_{\rm H}$$^{\rm a}$	& cos$i$$^{\rm b}$	& $T_{\rm s}$$^{\rm c}$	& $Z$$^{\rm d}$ 	& EW$^{\rm e}$	& EW$^{\rm f}$		& $F_{\rm 15-50}$$^{\rm g}$			& $F_{\rm 0.01-100}$$^{\rm h}$ 	& $\chi^{2}/\nu$		\\
			& 10$^{22}$ cm$^{-2}$	& 				& keV				& $Z_{\odot}$		& eV				& eV				& erg cm$^{-2}$ s$^{-1}$				& erg cm$^{-2}$ s$^{-1}$		&  		\\
\hline
Outburst		& 124 $^{+14}_{-12}$	& 				& 19.7$^{+1.3}_{-1.0}$	& 				& --				& $192 \pm 13$	& 4.3$^{+0.6}_{-0.9} \times 10^{-10}$	& 3.6$^{+0.5}_{-0.8} \times 10^{-9}$	&		\\
			& 					& $<$0.22			&					& 0.10 $\pm$ 0.04	&				&				& 								& 							& 1.01	\\
Quiescence	& 10 $^{+3}_{-4}$		& 				& 36.2$^{+3.5}_{-3.2}$	& 				& $55^{+34}_{-6}$	& $52^{+34}_{-26}$	& 1.3$^{+0.1}_{-0.2} \times 10^{-11}$		& 5.5$^{+0.5}_{-0.9} \times 10^{-11}$	&	\\
\hline
\multicolumn{7}{l}{$^{\rm a}$ Column density of the single-column absorption.} \\ 
\multicolumn{7}{l}{$^{\rm b}$ Cosine of the inclination angle between the reflection surface and the observer's line-of-sight.} \\ 
\multicolumn{7}{l}{$^{\rm c}$ The highest temperature of the accretion column.} \\ 
\multicolumn{7}{l}{$^{\rm d}$ Abundance relative to Solar.} \\ 
\multicolumn{7}{l}{$^{\rm e}$ Equivalent width of the Fe-K line by the 5--50 keV wide band fit.} \\ 
\multicolumn{7}{l}{$^{\rm f}$ Equivalent width of the Fe-K line by the 5--9 keV narrow band fit.} \\ 
\multicolumn{7}{l}{$^{\rm g}$ Obtained model fluxes in the 15--50 keV band.} \\ 
\multicolumn{7}{l}{$^{\rm h}$ Extrapolated total X-ray fluxes in the 0.01--100 keV band.} \\ 
\end{tabular}
\end{center}  
\end{table*}

\begin{figure}
	\begin{center}
	\includegraphics[width=0.47\textwidth]{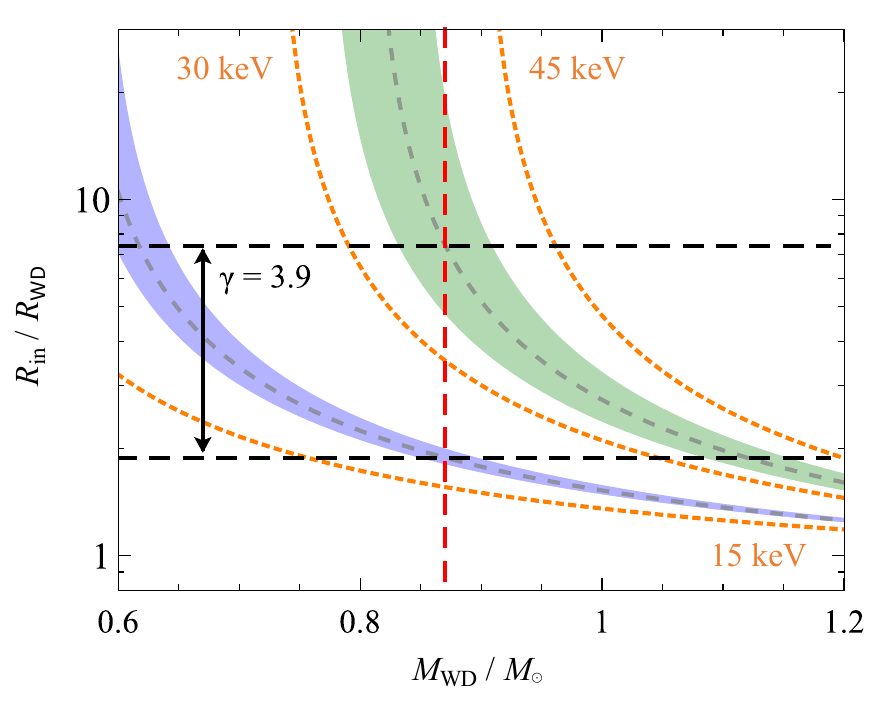}
	\caption{Contours (dotted orange curves) of $T_{\rm s}$ given by Equation~\ref{eq:suleimanov}, shown on the $M_{\rm WD}/M_{\rm \odot}$--$R_{\rm in}/R_{\rm WD}$ plane.
	Allowed regions for $T_{\rm s}$ in outburst and quiescence are indicated by blue and green, respectively. 
	The dashed red line (vertical) shows a value of $M_{\rm WD}$ that satisfies $\gamma = 3.9$ between the two measurements.}
	\label{fig:gamma-mass}
	\end{center}
\end{figure}

\section{Estimation of the WD mass}
\label{sec:estimation}
As described in Section~\ref{sec:introduction}, the accreting matter is considered to be captured by the magnetic field at the inner-radius of the accretion disk $R_{\rm in}$.
	In order to estimate the WD mass precisely, $R_{\rm in}$ as well as the shock height $h$, has to be taken into account. Thus, Equation~\ref{eq:aizu} is modified as
	\begin{equation}
		kT_{\rm s} = \frac{3}{8} \mu m_{\rm p} \frac{GM_{\rm WD}}{R_{\rm WD}} \left( \frac{R_{\rm WD}}{R_{\rm WD} + h}-\frac{R_{\rm WD}}{R_{\rm in}} \right),
		\label{eq:modified_aizu}
	\end{equation}
	and therefore $T_{\rm s}$ is a function of $M_{\rm WD}$, $R_{\rm WD}$, $h$, and $R_{\rm in}$.
	At the same time, $R_{\rm WD}$ is related to $M_{\rm WD}$ by the theoretical $M$--$R$ relation (Equation~\ref{eq:nauenberg}). 
	Employing a numerical calculation of plasma emission in the accretion column,
	\citet{2008A&A...491..525S,2016A&A...591A..35S} theoretically calculated the behaviour of $h$ and $T_{\rm s}$, and then reconstructed Equation~\ref{eq:modified_aizu} as the functional form, 
	\begin{equation}
		\frac{kT_{\rm s}}{\rm 1~keV} \approx 23.4 \frac{M_{\rm WD}}{M_{\rm \odot}} \left (1 - 0.59 \frac{M_{\rm WD}}{M_{\rm \odot}} \right)^{-1} \left(1- \frac{R_{\rm WD}}{R_{\rm in}} \right).
		\label{eq:suleimanov}
	\end{equation}
	Based on the article, this expression is valid for $M_{\rm WD}$ in the range of 0.4 $M_{\rm \odot}$ to 1.2 $M_{\rm \odot}$.
	Figure~\ref{fig:gamma-mass} presents, on the $M_{\rm WD}$ v.s. $R_{\rm in}$ plane, contours of $T_{\rm s}$ implied by Equation~\ref{eq:suleimanov}; 
	in this visualisation, we employed several representative values of $T_{\rm s}$, including the present two measurements for outburst and quiescence.
	Thus, from the quiescence data, we can already set a lower limit as $M_{\rm WD}~>~0.77~M_{\odot}$ (for $R_{\rm in} \to \infty$).

The inner-disk radius may be approximately identified as the Alfven radius, which is determined by an equilibrium between the inward gravity and the outward magnetic pressure.
	Assuming spherical accretion and dipole magnetic field, \citet{1977ApJ...215..897E} described it as
	\begin{equation}
		\frac{R_{\rm in}}{R_{\rm WD}} \approx 2.3 \: \left(\frac{\dot{M}}{10^{20} \: {\rm g/s}}\right)^{-2/7} \left( \frac{M_{\rm WD}}{M_{\rm \odot}} \right)^{-1/7} \:
		\left( \frac{R_{\rm WD}}{10^{9} \: {\rm cm}} \right)^{5/7}  \left( \frac{B}{10^{6} \: {\rm G}} \right)^{4/7},
		\label{eq:ghosh_lamb}
	\end{equation}
	where $\dot{M}$ is the accretion rate and $B$ is again the dipole magnetic-filed strength on the WD surface. 
	An extension of this formalism by \citet{1979ApJ...234..296G} has been shown to give a good explanation to the accretion-induced spin period changes
	in the binary X-ray pulsar 4U~1626--67 \citep{2016PASJ...68S..13T}. 
	Since mass accretion that takes place in GK Persei shares similar geometry outside $R_{\rm in}$, this formalism is employed here.
	With this formalisation, $R_{\rm in}$ thus shrinks as $\dot{M}$ increases, $T_{s}$ must be lower in outbursts than in quiescence, in agreement with our result.
	Systematic uncertainties associated with this equation are discussed in Section~\ref{sec:alfven-radius}.
	
In order to utilise our two observations in equal manner, let us introduce the ratio $\gamma$ between the two $R_{\rm in}$ values as
	\begin{equation}
		\gamma = \frac{R_{\rm in}^{\rm qui}}{R_{\rm in}^{\rm out}} = \left( \frac{\dot{M}^{\rm out}}{\dot{M}^{\rm qui}} \right)^{-2/7},
		\label{eq:gamma}
	\end{equation}
	where the superscripts ``out'' and ``qui'' indicate the outburst and quiescence values, respectively.
	Since $\dot{M}$ can be related to the total X-ray luminosity $L$ as
	\begin{equation}
		\dot{M} = L \times \left[ \frac{GM_{\rm WD}}{R_{\rm WD}} \left( 1-\frac{R_{\rm WD}}{R_{\rm in}} \right) \right]^{-1},
		\label{eq:accretion_rate}
	\end{equation}
	we can combine Equation~\ref{eq:suleimanov},~\ref{eq:gamma}, and~\ref{eq:accretion_rate} to obtain
	\begin{equation}
		\gamma = \left( \frac{L^{\rm qui} \times T_{\rm s}^{\rm out}}{L^{\rm out} \times T_{\rm s}^{\rm qui}} \right)^{-2/7}  = \left( \frac{F^{\rm qui} \times T_{\rm s}^{\rm out}}{F^{\rm out} \times T_{\rm s}^{\rm qui}} \right)^{-2/7},
		\label{eq:gamma_equated}
	\end{equation}
	where $F$ is the total X-ray flux. Then, the values of $T_{\rm s}$ and $F$ measured in the two observations (Table~\ref{table:parameter}) yield, via Equation~\ref{eq:gamma},
	\begin{equation}
		\gamma = 3.9 \pm 0.5.
		\label{eq:gamma_value}
	\end{equation}
	
We can now determine the value of $M_{\rm WD}$ in Figure~\ref{fig:gamma-mass} so that the two $R_{\rm in}$ values satisfy equation~\ref{eq:gamma_value} as indicated by a pair of horizontal lines
	in Figure~\ref{fig:gamma-mass}; $R_{\rm in}^{\rm out} = 1.9^{+0.4}_{-0.2}~R_{\rm WD}$, $R_{\rm in}^{\rm qui} = 7.4^{+2.1}_{-1.2}~R_{\rm WD}$, 
	and $M_{\rm WD} = 0.87~\pm~0.05~M_{\rm \odot}$ are yielded with $R_{\rm WD} = (6.6 \pm 0.4) \times 10^{8}~{\rm cm}$.
	The obtained $R_{\rm in}^{\rm qui}$ and $R_{\rm in}^{\rm out}$ satisfy the accretion condition $R_{\rm in} < R_{\rm \Omega} \sim 11~R_{\rm WD}$
	with $P=351~{\rm sec}$, $M_{\rm WD} = 0.87~M_{\rm \odot}$, and Equation~\ref{eq:corotation}.
	Thus, the clear increase in $T_{\rm s}$, observed in the transition from outburst to quiescence  (Table~\ref{table:parameter}), has been successfully explained as a factor $\gamma \sim 4$ increase in $R_{\rm in}$, 
	in response to the decrease in $\dot{M}$ by a factor of 120: 
	$\dot{M}$ is obtained from Equation~\ref{eq:accretion_rate} and $F$ as $\dot{M}^{\rm out} = (1.2 \pm 0.4)\times10^{18}~{\rm g~s^{-1}}$
	and $\dot{M}^{\rm qui}~=~(1.0 ~ \pm ~ 0.3) \times 10^{16}~{\rm g~s^{-1}}$, respectively.
	An advantage of this method is that any systematic uncertainty involved in the coefficient of Equation~\ref{eq:ghosh_lamb} cancels by taking the ratio of the two equations (Equation~\ref{eq:gamma_equated}).
	Further discussion continues in Section~\ref{sec:alfven-radius}.

\section{Discussion}
\label{sec:discussion}
\subsection{Comparison between the Two Observations}
\label{sec:comparison1}
In the present paper, we analysed a pair of {\sl NuSTAR} spectra of GK Persei acquired in 2015.
	In the 5 months from the outburst observation on April 4 to the quiescence one  on September 8,
	the optical emission from GK Persei diminished by $\sim$3.2 magnitude or $\sim$19 times (Figure~\ref{fig:monitor}).
	Meanwhile, the 3--50 keV FPMA+FPMB count rate decreased by a factor of 17, and the absorbed 15--50 keV  flux by a factor of 33.
	The two factors become different because the outburst spectrum is more absorbed (Figure~\ref{fig:ratio}), and hence the count rate which is more weighted towards lower energies 
	changed less than that of the flux which is more weighted towards higher energies.
	Correcting these spectra for the respective absorption, and extrapolating the best-fit models to $>$50~keV and $<$15~keV, the 0.01--100 keV unabsorbed total X-ray flux is inferred to have changed by 65 times.

In addition to the changes in the X-ray flux and absorption, we detected a clear increase in $T_{\rm s}$ from the outburst ($\sim 20$ keV) to the quiescence ($\sim 36$ keV) observations.
	Employing the disk-magnetosphere interaction model of \citet{1979ApJ...234..296G}, the change in $T_{\rm s}$ has successfully been interpreted
	as due to a factor $\sim 4$ change in $R_{\rm in}$, in a negative correlation with $\dot{M}$.
	Considering that the gravitational potential drop available for the X-ray emission (from $R_{\rm in}$ to $R_{\rm WD}$) thus became deeper in quiescence,
	the total X-ray luminosity change has been converted to a factor of 120 difference in $\dot{M}$.
	For reference, the temperature change we observed is qualitatively consistent with the report by \citet{2017MNRAS.469..476Z}, 
	that the {\sl Swift}/XRT light curve of the hardness ratio indicated a temperature decrease as the outburst proceeded towards its peak,
	and the very high value of $T_{\rm s}$ measured by \citet{2016MNRAS.459..779Y} at the outburst onset.

As presented in Table~\ref{table:parameter}, the EW of the Fe-K line (nearly neutral component) was $\sim$~4~times higher in outburst. 
	The line is usually ascribed to two emitting sources: the ambient matter as represented by $\dot{M}$ and $N_{\rm H}$, and the WD surface as represented by reflection.
	Evidently, both $\dot{M}$ and $N_{\rm H}$ were higher in outburst, so that the Fe-K line EW from the first source must be higher as well.
	Furthermore, as discussed later in Section~\ref{sec:reflection}, the standing shock is considered to come closer to the surface when $\dot{M}$ increases, 
	because the higher density would increase the volume emissivity in the accretion columns and higher shock temperature can be dissipated within the boundary conditions.
	This will in turn increase the solid angle of reflection, and yield a high EW from the second source.
	The higher EW observed in outburst may be explained qualitatively as a combination of these two effects.
	
\subsection{Comparison with Previous Optical Results}
\label{sec:comparison2}
The WD mass we obtained, $M_{\rm WD} = 0.87 \pm 0.05 ~M_{\rm \odot}$, is consistent with the optical results
	($\geq 0.78~M_{\rm \odot}$: \citealt{1994A&A...281..108R}, $\geq 0.63~M_{\rm \odot}$: \citealt{2002MNRAS.329..597M}).
	These optical estimates gave only lower limits of $M_{\rm WD}$ because eclipses of the WD do not occur in the GK Persei system,
	and hence the inclination angle remains poorly constrained.
	In contrast, our method with X-rays can estimate $M_{\rm WD}$ without the knowledge of the inclination angle.
	When our $M_{\rm WD}$ determination is combined with the ratio of the WD mass and the companion mass $M_{\rm K}/M_{\rm WD} = 0.55 \pm 0.21$, and the optically determined mass function
	\begin{equation}
		\frac{M_{\rm WD}}{M_{\rm \odot}}\frac{\sin^{3}i}{(1+M_{\rm K}/M_{\rm WD})^{2}} = 0.362
	\end{equation}
	\citep{2002MNRAS.329..597M}, the companion mass is constrained as $M_{\rm K} = 0.48 \pm 0.18~M_{\rm \odot}$, and then the lower limit of the inclination angle is derived as $i \geq 63^{\rm \circ}$.
	This lower limit on $i$ is consistent with the optical upper limit, $i \leq 73^{\rm \circ}$, required by the lack of eclipses.
	Combining these results, $63^{\rm \circ} \leq i \leq 73^{\rm \circ}$ is obtained.

\subsection{Comparison with Previous X-ray Results}
\label{sec:comparison3}
Let us revisit the past X-ray result with the PCA and HEXTE onboard {\sl RXTE}, obtained during an outburst by \citet{2005A&A...435..191S}. 
	They measured $T_{\rm s}=21\pm3~{\rm keV}$, and derived $M_{\rm WD}=0.59 \pm 0.05~M_{\rm \odot}$ assuming $R_{\rm in} \gg R_{\rm WD}$
	(the author noted that $M_{\rm WD}$ would be underestimated).
	At that time, the total X-ray flux in the 0.1--100.0~keV range was measured to be $8.86 \times 10^{-10}$~erg~cm$^{-2}$~s$^{-1}$.
	When we use the present {\tt cemekl} model of the same $T_{\rm s}$, the 0.01-100~keV flux is re-estimated as $1.0 \times 10^{-9}$~erg~cm$^{-2}$~s$^{-1}$,
	which falls in between the present two measurements (Table~\ref{table:parameter}).
	Then, compared with them, the value of $R_{\rm in}$ during the {\sl RXTE} observation is estimated,
	from Equation~\ref{eq:gamma}, as $R_{\rm in} \simeq 1.5~R_{\rm in}^{\rm out} \simeq 0.37~R_{\rm in}^{\rm qui} \simeq 2.8~R_{\rm WD}$.
	Substituting this value and $T_{\rm s} = 21~{\rm keV}$ into Equation~\ref{eq:suleimanov}, or equivalently referring to Figure~\ref{fig:gamma-mass},
	$M_{\rm WD} \sim 0.8~M_{\rm \odot}$ is derived. This revised mass is probably consistent with our result when various errors are taken in account.
	Our result is also consistent with the mass estimation $M_{\rm WD} = 0.90 \pm 0.12~M_{\rm \odot}$ by \citet{2009A&A...496..121B} with {\sl Swift}/BAT within errors,
	as already referred to in Section~\ref{sec:introduction}.

The present outburst data were already analysed by \citet{2017MNRAS.469..476Z}.
	They derived $T_{\rm s} = 16.2^{+0.5}_{-0.4}~{\rm keV}$ by the 3--50~keV broad band fitting, wherein the {\sl NuSTAR} data are combined with those from the {\sl Chandra} MEG and HETG.
	Their $T_{\rm s}$ value is $\sim$18\% lower than our outburst result. This discrepancy may be caused by difference of emission models.
	The {\tt mkcflow} model they employed is a superposition of the {\tt mekal} thermal emission model, like the {\tt cemekl} model we used,
	but the emission measure of {\tt mkcflow} is weighted by the inverse of the bolometric luminosity at each temperature $T$.
	Because the bolometric flux is $\propto T^{1/2}$ when only the bremsstrahlung continuum is considered,
	the differential emission measure becomes $\propto (T/T_{\rm s})^{-1/2}$, and $\alpha = 0.5$ by Equation~\ref{eq:emission}.	
	It is slightly different from that of {\tt cemekl}, $\alpha = 0.43$, and will make the composite spectrum more weighted towards higher temperatures.
	In the fit by \citet{2017MNRAS.469..476Z}, this effect is considered to be compensated by the lower $T_{\rm s}$. 
	(In the relevant temperature range, $\alpha$ would not change very much even considering the lines.)

\citet{2016A&A...591A..35S} also analysed the same outburst data and obtained $M_{\rm WD} = 0.86 \pm 0.02~M_{\rm WD}$. This is fully consistent with our estimate.
	In deriving this result, however, they employed a method that differs from ours in two points: $T_{\rm s}$ and $R_{\rm in}$.
	They fitted the 20--70 keV spectrum in outburst with their newly calculated spectral model (PSR model), and obtained $T_{\rm s} \sim 26.3$~keV.
	They also employed, for an illustrative purpose, a single temperature bremsstrahlung model and obtained its temperature as $16.7 \pm 0.2$~keV;
	via Equation~2 in  \citet{2016A&A...591A..35S}, this was converted to a consistent shock temperature of $T_{\rm s} = 26.0 \pm 0.3$~keV.	
	For consistency, we fitted the 20--70 keV spectrum with the bremsstrahlung model, and obtained the temperature as $16.6 \pm 0.3$~keV. Therefore, the present data analysis is consistent with theirs.
	They also derived $R_{\rm in} = 2.8 \pm 0.2~R_{\rm WD}$ by the power density spectral analysis \citep{2009A&A...507.1211R}.
	After all, their $T_{\rm s}$ is 1.3 times higher than our $T_{\rm s}$, and their $R_{\rm in}$ is 1.5 times larger than ours.
	These differences in $T_{\rm s}$ and $R_{\rm in}$ happened to cancel out, to yield the two $M_{\rm WD}$ estimates which are very close to each other.
	
\subsection{Systematic Uncertainties of the Mass Estimation}
\label{sec:uncertainty}
So far, we considered only statistic errors. Here, let us evaluate possible systematic errors that can affect our result.

\subsubsection{Emission Models of the Accretion Column}
\label{sec:model}
In the present paper, we assumed the accretion columns to have a cylindrical shape, and hence employed the {\tt cemekl} model with $\alpha=0.43$ in Equation~\ref{eq:emission}.
	Recently, emission models with a dipole geometry for the accretion column have been developed,
	including ACRAD model by \citet{2014MNRAS.438.2267H,2014MNRAS.441.3718H}, and the PSR model by \citet{2016A&A...591A..35S}. 
	We thus refitted the spectra with the PSR model ({\tt ipolar} model in XSPEC).
	The energy range below 7~keV in quiescence was ignored in this analysis because {\tt ipolor} model has abundances fixed to $1~Z_{\rm \odot}$
	and cannot reproduce the Fe emission lines which are well described with sub-solar Fe abundance.
	The shock temperature was then obtained as $20.4 \pm 0.6$~keV in outburst and $37.3 ^{+3.9}_{-3.3}$~keV in quiescence, 
	and the WD mass was constrained as $M_{\rm WD} = 0.88 \pm 0.05 M_{\rm \odot}$.
	All these values agree well with the results in Section~\ref{sec:result} within the statistic errors.
	Therefore, we consider that slight differences of the emission models, namely the detailed morphological and emissivity structures of the post-shock region, 
	have insignificant impact on the mass estimation method presented above, at least, when applied to GK Persei.
	
\subsubsection{Shock Height}
\label{sec:reflection}
As presented in Equation~\ref{eq:modified_aizu}, $T_{\rm s}$ depends on the shock height $h$, which is thought to negatively correlate with $\dot{M}$.
	This effect was theoretically calculated and incorporated in Equation~\ref{eq:suleimanov}.
	However, even if the effect of $h$ is ignored (i.e. $h=0$), the value of $M_{\rm WD}$ changes by less than 0.5\%, which is much smaller than the statistic error.
	Therefore, the value of $h$ itself would not affect the WD mass estimation in GK Persei.
	
In our spectral analysis, the reflection model was included to represent the reflection effect on the WD surface.
	The solid angle of the reflection was fixed to $2\pi$ simply assuming that the shock heating occurs just above the WD surface.
	In reality, $h$ are calculated as 0.014~$R_{\rm WD}$ and 0.026~$R_{\rm WD}$ by Equation~\ref{eq:modified_aizu},
	corresponding to the solid angle of 1.8$\pi$ and 1.7$\pi$ in outburst and quiescence, respectively.
	We thus repeated the spectral fitting using these solid angles, to find that neither $T_{\rm s}^{\rm out}$ nor $T_{\rm s}^{\rm qui}$ changes by more than 0.2\%
	Therefore, the result of $M_{\rm WD}$ is not affected either.

\subsubsection{Alfven Radius}
\label{sec:alfven-radius}
As expressed by Equation~\ref{eq:ghosh_lamb}, we assumed that $R_{\rm in}$ is equal to the Alfven radius.
	This formalism by \citet{1977ApJ...215..897E}, which considers spherical accretion, includes two possible uncertainties.
	One is the coefficient of Equation~\ref{eq:ghosh_lamb}. In the case of disk accretion in rotating magnetic neutron star systems, 
	\citet{1979ApJ...234..296G} argued that the actual inner-disk radius is almost half the Alfven radius.
	Therefore, the estimated values of $R_{\rm in}$ are probably subject to an uncertainty by a constant factor.
	However, this uncertainty cancels out in our work by taking the ratio $\gamma = R_{\rm in}^{\rm qui}/R_{\rm in}^{\rm out}$.
	Hence, our $M_{\rm WD}$ estimation is free from the uncertainty, because in Figure~\ref{fig:gamma-mass} we utilise $\gamma$ rather than the actual values of $R_{\rm in}$.
	
The other uncertainty is the power law index $\Gamma$ of $\dot{M}$ in Equations~\ref{eq:ghosh_lamb}, \ref{eq:gamma}, and \ref{eq:gamma_equated}, for which we employ $\Gamma = -2/7$.
	A recent 3D simulation of accretion flows around a neutron star yields $\Gamma = -1/5$ \citep{2013MNRAS.433.3048K}. 
	In addition, \citet{2016A&A...591A..35S} observationally evaluated $\Gamma = -0.2^{+1.0}_{-1.5}$.
	We thus quote a systematic uncertainty by $\Delta \Gamma \sim 0.1$, which translates into $\sim$~7\% systematic errors in $M_{\rm WD}$.

Adding up all these uncertainties, the overall systematic error in the WD mass estimate becomes comparable to the statistical error.
	Including this in quadrature, we quote our final mass determination as $M_{\rm WD}=0.87 \pm 0.08~M_{\rm \odot}$, with $R_{\rm WD}=(6.6 \pm 0.6) \times 10^{8}~{\rm cm}$.

\subsection{Strength of Magnetic Field on WD Surface}
\label{sec:magnetic}
Since $M_{\rm WD}$, $R_{\rm WD}$, and $R_{\rm in}$ were determined, the strength of the magnetic field on the WD surface can be now estimated to be $B \sim 5 \times 10^{5} \: {\rm G}$ 
	using Equation~\ref{eq:ghosh_lamb} and the distance of the object 477~pc \citep{2013ApJ...767....7H}.
	This value is consistent with the typical magnetic field strength of IPs. However, unlike $M_{\rm WD}$, this result is subject to the uncertainties discussed in Section~\ref{sec:alfven-radius}.
	If the coefficient of Equation~\ref{eq:ghosh_lamb} has an uncertainty by a factor of 2 for example, $B$ changes by a factor of 3--4.

Magnetic field measurements of Polars and strongly magnetised IPs ($\sim 10^{7}$~G) have been made 
	by detecting spin-modulated polarisation in the near-ultraviolet to near-infrared bands (e.g. \citealt{2008ApJ...684..558P}). 
	However, the magnetic field of $\sim 10^{5}$~G on the WD surface cannot be measured at present. 
	Therefore, the present method provides the only way to measure relatively weak magnetic field of IPs, even though it has a relatively poor accuracy due to the above-mentioned model uncertainty.

\section{Conclusion}
Analysing the outburst and quiescence data of GK Per obtained with {\sl NuSTAR}, 
	we found the 0.01--100 keV unabsorbed flux was $3.6^{+0.5}_{-0.8} \times 10^{-9}~{\rm erg}~{\rm s}^{-1}~{\rm cm}^{-2}$ and 5.5$^{+0.5}_{-0.9} \times 10^{-11}~{\rm erg}~{\rm s}^{-1}~{\rm cm}^{-2}$, 
	respectively, with a factor 65 difference.
	Analysing the 5--50 keV or 15--50 keV spectra using a multi-temperature spectral model, 
	the shock temperature was determined as $19.7^{+1.3}_{-1.0}$~keV in outburst, and $36.2^{+3.5}_{-3.2}$~keV in quiescence.
	Assuming that this temperature difference is caused by a compression of the magnetosphere and the associated change in $R_{\rm in}$,
	we determined the WD mass in GK Per as $M_{\rm WD} = 0.87 \pm 0.08~M_{\rm \odot}$, together with the radius as $R_{\rm WD}=(6.6 \pm 0.6) \times 10^{8}~{\rm cm}$ (including a 7\% systematic error).
	The values of $R_{\rm in}$, relative to $R_{\rm WD}$, was derived as $R_{\rm in}/R_{\rm WD}=1.9^{+0.4}_{-0.2}$ in outburst
	and $R_{\rm in}/R_{\rm WD}=7.4^{+2.1}_{-1.2}$ in quiescence, and the mass accretion rate is estimated to have changed by a factor of 120 between the two observations.
	Combined with optical observations, the inclination angle of GK Per is tightly constrained as $63^{\circ} \leq i \leq 73^{\circ}$.
	We also estimated the magnetic field of the WD as $\sim 5 \times 10^{5}$~G although it is subject to large systematic error uncertainty.
	The overall results demonstrate the power of our mass determination method using X-ray luminosity changes,
	wherein some major systematic uncertainties cancel out.

\section*{Acknowledgements}
This work was made use of data obtained with the {\sl NuSTAR} mission, and softwares obtained from the High Energy Astrophysics Science Archive Research Center at NASA Goddard Space Center.
	The authors also acknowledge the use of X-ray monitoring data provided by {\sl Swift} data archive, and optical data from {\sl AAVSO International Database} contributed by observers worldwide.
	Y.W. is supported by the Junior Research Associate Programme in RIKEN.




\bibliographystyle{mnras}
\bibliography{gkper_ref}








\bsp	
\label{lastpage}
\end{document}